# Avoiding Register Overflow in the Bakery Algorithm

Amirhossein Sayyadabdi

Faculty of Computer Engineering, University of Isfahan, Isfahan, Isfahan, Iran, ahsa@eng.ui.ac.ir

Mohsen Sharifi

School of Computer Engineering, Iran University of Science and Technology, Tehran, Tehran, Iran, msharifi@iust.ac.ir

## ABSTRACT

Computer systems are designed to make resources available to users and users may be interested in some resources more than others, therefore, a coordination scheme is required to satisfy the users' requirements. This scheme may implement certain policies such as "never allocate more than X units of resource Z". One policy that is of particular interest is the inability of users to access a single resource at the same time, which is called the problem of mutual exclusion. Resource management concerns the coordination and collaboration of users, and it is usually based on making a decision. In the case of mutual exclusion, that decision is about granting access to a resource. Therefore, mutual exclusion is useful for supporting resource access management. The first true solution to the mutual exclusion problem is known as the Bakery algorithm that does not rely on any lower-lever mutual exclusion. We examine the problem of register overflow in real-world implementations of the Bakery algorithm and present a variant algorithm named Bakery++ that prevents overflows from ever happening. Bakery++ avoids overflows without allowing a process to write into other processes' memory and without using additional memory or complex arithmetic or redefining the operations and functions used in Bakery. Bakery++ is almost as simple as Bakery and it is straightforward to implement in real systems. With Bakery++, there is no reason to keep implementing Bakery in real computers because Bakery++ eliminates the probability of overflows and hence it is more practical than Bakery. Previous approaches to circumvent the problem of register overflow included introducing new variables or redefining the operations or functions used in the original Bakery algorithm, while Bakery++ avoids overflows by using simple conditional statements. The result is a new mutual exclusion algorithm that is guaranteed never to allow an overflow and it is simple, correct and easy to implement. Bakery++ has the same temporal and spatial complexities as the original Bakery. We have specified Bakery++ in PlusCal and we have used the TLC model checker to assert that Bakery++ maintains the mutual exclusion property and that it never allows an overflow.

## 1 Introduction

Concurrent computing is a world with its own physics that is very similar to our world, wherein independent entities live their separate lives and communicate with each other to achieve certain goals. A major goal of a concurrent computer system is to make resources available. Computer resources cover a wide range of different types of goods such as registers, central processing unit (CPU) cycles, printers, files and execution of instructions.

Where there is a scarcity of a resource, it is rational to share that resource among those who need it. The most common example is a CPU that is shared among multiple processes. Resource sharing is a procedure that requires the collaboration of system participants. According to the system model presented in [3], a system consists of a group of cyclic processes, meaning that they repeatedly execute a finite set of instructions over and over. Processes communicate by reading and writing shared units of memory. Whenever there are multiple read or write requests for a memory location, operations are executed one at a time in an arbitrary order. Each process has a portion of instructions that is called its *critical section*.

## 1.1 Mutual Exclusion

The problem of mutual exclusion is concerned with devising a protocol for processes that guarantees no two processes will ever execute their critical sections at the same time. In other words, if a process is executing its critical section then no other process is allowed to execute its critical section. The original version of the problem, introduced and solved by Dijkstra [3], requires five conditions to hold for a correct solution:

- No two processes are allowed to execute their critical sections simultaneously.
- The solution must treat all processes equally. As an exemplar, assigning priorities to processes is not allowed.
- No assumption is to be made about the execution speeds of processes.
- If a process crashes (halts) outside of its critical section, then it should not cause other processes who wish to enter their critical sections to be blocked forever.
- If there are some processes that wish to enter their critical sections, then the decision of which one is to be allowed should be made in a finite length of time.

The problem of mutual exclusion is foundational in concurrent computing and the reason may be the problem of consensus[1] that lies at the heart of mutual exclusion itself. The ability to have mutually exclusive access to a resource can provide processes with countless applications. Two threads can have concurrent access to a shared data structure. A group of users may share the same file and make sure that no two users ever edit the file at the same time. Lock services may be implemented by a group of computers that wish to provide locking mechanisms for any shared resource such as a database table.

One important application of mutual exclusion is in the context of resource access management in parallel systems. Mutual exclusion algorithms do not allow parallel processes to access a specified resource at the same time; therefore, they facilitate the management of access to resources.

## 1.2 The Bakery Algorithm

The first solution to the problem of mutual exclusion was proposed and proved correct by Dijkstra [3] in 1965, and that was not the end of the story. Nine years later, a new algorithm named *Bakery* was introduced by Lamport [6] that has four remarkable properties:

- Processes can enter their critical sections in first-come-first-served order.
- The failure of individual system components will not cause the entire system to halt.
- No process writes into the memory of other processes.
- For any memory location, if a read operation occurs simultaneously with a write operation, then the value obtained by the read operation may have any arbitrary value.

The Bakery algorithm assumes the same system configuration as the original paper by Dijkstra did. Knuth [5] extended the conditions of a correct solution that was originally proposed by Dijkstra, and Lamport [6] used it for the Bakery algorithm as well:

1. No two processes are allowed to execute their critical sections simultaneously.
2. If a process does not crash, then it should be allowed to enter its critical section in a finite length of time.

---

[1] The problem of consensus is concerned with collaboration on reaching an agreement over a subject.

3. Processes are allowed to crash outside their critical sections, and other processes should not be blocked from accessing their critical sections.
4. A process may fail at any instant but it can restart outside its critical section by resetting the shared memory units that belong to itself equal to the initial values. Specifically, if a process crashes, then it goes to its noncritical section and halts and any read operation from its memory units is expected to return 0 eventually.
5. No assumptions are to be made about the execution speeds of processes.

## 1.3 Trouble with Bakery in Real Systems

The Bakery algorithm, once implemented in real systems, has a drawback that relates to the overflow of registers because Bakery assumes unbounded registers. In this paper, we present a variant of Bakery, named Bakery++, which avoids overflows. Previous approaches to avoiding overflows included introducing new shared variables or redefining certain operations or functions in the Bakery algorithm. Bakery++ avoids overflows by simple conditional statements. The result is a new algorithm that is almost as easy as Bakery to implement and yet it never allows an overflow in real systems, which is of interest to practical systems. Bakery++ can be easily deployed to support resource access management in parallel systems without register overflows occurring.

The mutual exclusion problem is fundamental in computer resource management, and the Bakery algorithm has proved to be useful in application scenarios that limit the number of concurrent access to a resource to 1. Bakery++ tries to overcome the problem of register overflow that will occur in real-world implementations of the Bakery algorithm; therefore, Bakery++ is practical for resource management in parallel and distributed systems.

We have structured the rest of the paper as follows. Section 2 gives a background of the Bakery algorithm. Section 3 discusses the problem of register overflow in real-world implementations of Bakery. Section 4 explains previous approaches to overcoming the problem of register overflow. Section 5 introduces a new variant of Bakery named Bakery++ that avoids overflows. In Section 6 we prove that Bakery++ never allows an overflow and argue that Bakery++ is a refinement of the Bakery algorithm. Section 7 discusses the temporal and spatial complexities of Bakery++ and its practicality. Section 8 expresses two open questions concerning the newly introduced Bakery++ and Section 9 concludes the paper.

## 2 Background

Bakery is the first *true* solution to the problem of mutual exclusion because it does not rely on *any* lower-level mutual exclusion, such as atomic read and write operations [8]. Imagine a group of customers in a bakery all waiting to be served hot bread. There is a teller machine in the bakery that prints natural numbers on a ticket for customers. Each customer will receive a ticket with a number on it, and the customer with the lowest-numbered ticket is the one to be served. A customer is modeled as a process and the procedure for each process is depicted in Algorithm 1.

Algorithm 1. Original Bakery for N Processes [6]

**integer array** choosing [1..N], number [1..N]; \* Declaring shared variables with initial values equal to 0.

**begin integer** j; \* Procedure for process numbered *i*

    L1: choosing[i] := 1;

    \* the maximum function can take its argument in any arbitrary order.

    number[i] := 1 + maximum (number[1], …, number[N]);

    choosing[i] := 0;

```
        for j = 1 step 1 until N do

        begin

        L2: if choosing[j] ≠ 0 then goto L2;

        \* The operator "<" takes two operands that are ordered pairs of integers, namely (a, b)

        \* and (c, d). It returns true if (a < c) or ((a = c) and (b < d)).

        L3: if number[j] ≠ 0 and (number[j], j < number[i], i) then goto L3;

        end;

        critical section;   number[i] := 0;   noncritical section;         goto L1;

end
```

## 3  Problem Statement

The Bakery algorithm assumes registers that can hold arbitrary large values, meaning that there is no upper limit on the value that can be stored in variable number[i] for process *i*. According to [6], if there is always at least one customer in the bakery, then the natural number stored in the register named number can become arbitrarily large. This is not a problem at all in the world of algorithms, because algorithms are more abstract than programs. Consequently, an algorithm can handle natural numbers but a program cannot. The reason is that programs are executed by computers that have limited memory and storage, therefore, there is an upper limit on the maximum value a computer can handle.

We have specified the Bakery algorithm in pseudo-code in Algorithm 1. What happens next? A system builder will try to implement this algorithm in a real-world system. The person in charge of this task will need to take certain measures to handle the possibility of register values tending to infinity. Real computer registers are finite and there is always the possibility of an overflow that will cause Bakery to malfunction.

Consider a computer C that will be used to execute an implementation of the Bakery algorithm. C has a collection of registers that will store the values for variables number, choosing and j. No register can store a value $v \geq M$. We say an overflow occurs if C tries to store a value $v > M$ in one of its registers.

The Bakery algorithm has the potential of causing register overflows. Consider the following scenario in which two processes compete for entering their critical sections. Processes *A* and *B* are coordinating their actions by executing Bakery. Both processes are initially idle until *A* decides to enter its critical section by choosing ticket value 1. *A* successfully enters its critical section and in the meantime, *B* also decides to enter its critical section by choosing ticket value 2. Afterward, both *A* and *B* enter their critical sections exactly one after the other. Observe that the ticket value keeps increasing and tends to infinity.

It is important to note that algorithms that assume atomic read/write operations are not true mutual exclusion algorithms, because they assume lower-level mutual exclusion. We are interested in solutions that eliminate the possibility of overflow not those that have a small chance of overflow in practice.

It may seem that specific memory and cache models render Bakery++ unusable, but that discussion is irrelevant because the algorithm is abstract and its memory locations can be emulated by other means such as files. Besides, any modern general purpose CPU can be expected to dedicate two registers to the process for the purposes of Bakery++.

## 4  Related Work

The Bakery algorithm is the first true solution to the problem of mutual exclusion because it does not assume any lower-level mutual exclusion, and much work has been done to study and extend the algorithm's remarkable properties. Several works concentrate on modifying Bakery to avoid overflows. The majority of approaches include:

1. Changing the definitions of "<" operator and "maximum" function and using modulo arithmetic instead of integer arithmetic.
2. Introducing new shared variables or using extra memory.
3. Resetting the values of registers before an overflow occurs.

Jayanti et al. [4] followed approaches 1 and 2. Taubenfeld [11] followed the second approach. We will use the third approach in Bakery++, similar to the path taken by Yoon et al. [12] to avoid overflows in their distributed lock manager but their scheme is complicated. Aravind [1] introduced the symmetric token Bakery algorithm that follows the second approach and Aravind et al. [2] used the second approach to introduce the non-atomic dual Bakery algorithm.

The curious reader may point out that if we use 64-bit processors then the problem of register overflow would be extremely unlikely to happen. The probability of an overflow incident is related to several factors such as the processing speed of processes and the severity of competition between them. It is possible that Bakery may malfunction due to integer overflow in a 32-bit processor in less than a minute [1]. Many modern embedded systems are 32-bit machines for several reasons such as power consumption. To answer the curious reader's claim above, we ask another question: why use Bakery and worry about overflow while you can use Bakery++ and avoid it? Bakery++ is almost identical to Bakery and it is quite simple, which makes it a natural choice when trying to implement distributed shared-memory mutual exclusion algorithms. Bakery++ never allows an overflow by introducing simple conditional statements. Therefore, Bakery++ is fundamentally different from all existing solutions to the problem of register overflow in Bakery.

Szymanski's first-come-first-served algorithm [10] is much more complicated than Bakery++, and it also uses two more shared variables compared to Bakery++. Peterson's algorithm uses a variable "turn" that is shared by all processes, i.e. all processes are allowed to write into its memory location (as can be seen in the derivation of Peterson's algorithm presented in [9]). This separates our work from Peterson's algorithm because in Bakery and Bakery++, all processes write only into their own memory cells. In terms of complexity comparison between Bakery++ and Peterson's algorithm, Bakery++ uses more memory cells because it does not use a shared variable like "turn". Both algorithms have space complexity $O(n)$.

## 5 Bakery++

There is an important theoretical question in the paper that introduced Bakery [6]: "Can one find an algorithm for finite processors such that processors enter their critical sections on a first-come-first-served basis, and no processor may write into another processor's memory?"

To our knowledge, all of the previous works on bounding the Bakery algorithm have failed to answer this question. Jayanti et al. [4], Taubenfeld [11], Yoon et al. [12], Aravind [1] and Aravind et al. [2] have all failed to design their algorithms in such a way that no process writes into other processes' memory. In response to the question above, Lamport has stated in the same paper "we have recently found such an algorithm, but it is quite complicated". Our solution, named Bakery++, avoids overflows easily.

Bakery++ is a slightly modified version of the Bakery algorithm, presented in Algorithm 2. We call it Bakery++ because it is almost identical to Bakery. It does not use any additional variables and it does not redefine the operators or functions used in Bakery.

The basic idea in Bakery++ is driven by another statement by Lamport in [6] that says "*if there is always at least one processor in the bakery, then the value of number[i] can become arbitrarily large*". What Bakery++ does is that it emulates a system failure in which all processes have failed and restarted. Before Bakery++, every variant of Bakery that tried to avoid overflows either used more shared variables, or redefined operations or functions of the Bakery algorithm.

It may seem that label L1 at the beginning of the Bakery++ algorithm is an infinite waiting loop, but the instruction "number[i] := 0;" at the end of the algorithm ensures that all processes will reset their number[i] value. Therefore, any process with number[i] ≥ M will reset its value eventually. Bakery++ defines a constant M representing the maximum value storable in registers. Initially, Bakery++ checks if there exists a register that contains value ≥ M. If there is, then it waits for the owner of that register to reset it to 0. After that, the process chooses a maximum value among the registers. If the value is greater than or equal to M, then the process will reset number[i] and choosing[i] to 0 and jumps to the beginning of label L1. Else, the value is incremented and the rest of the operations are performed exactly like the original Bakery.

In Bakery++, if a process has variable number[i] equal to 0, that does not mean that it is not waiting to enter its critical section. The ability to detect a "waiting process" is not in question. It may seem that in the original Bakery algorithm a process with number[i] equal to 0 is a process that is not waiting, but that is not true because it depends on when a process is considered to be waiting. For example, a process that has just set its variable choosing[i] equal to 1 may be considered "waiting" as well! In Bakery++, a non-zero value for number[i] does not mean that process $i$ is has taken a ticket because that process has to check its value before incrementing it. The reason we used the operator $\geq$ is that Bakery assumes that a read that overlaps a write can return an arbitrary natural value. If we can assume that no value greater than the register limit M will ever be returned, then the operator = can also be used.

---

Algorithm 2. Bakery++ for N Processes

**constant** M; \* Declaring a constant for the maximum value of a register.

**integer array** choosing [1..N], number [1..N]; \* Declaring shared variables with initial values equal to 0.

**begin integer** j; \* Procedure for process numbered $i$

    L1: **if** $\exists$ q $\in$ {1, ..., N} such that number[q] $\geq$ M     **then goto** L1;

    choosing[i] := 1;

    \* the maximum function can take its argument in any arbitrary order.

    number[i] := maximum (number[1], ..., number[N]);

    **if** number[i] $\geq$ M

    **then begin**

        number[i] := 0;

        choosing[i] := 0;

        **goto** L1;

      **end**

    **else**     number[i] := number[i] + 1;

    choosing[i] := 0;

    **for** j = 1 **step** 1 **until** N **do**

    **begin**

    L2: **if** choosing[j] $\neq$ 0 **then goto** L2;

    L3: **if** number[j] $\neq$ 0 and (number[j], j < number[i], i) **then goto** L3;

    **end;**

    critical section;    number[i] := 0;    noncritical section;      **goto** L1;

**end**

# 6  Correctness Argument

Bakery++ is a modified version of the Bakery algorithm that does not introduce new variables and its control flow is almost identical to Bakery. The changes that have been made to Bakery++ include:

- Introducing a constant value that represents the maximum value that can be stored in a computer register.
- Adding a conditional statement and a ***goto*** after label L1 that does not manipulate the values of Bakery's data objects.
- Adding a conditional statement before incrementing the maximum value obtained from reading all processes' variable *number*. If there is a possibility of overflow in process *i*, then we simply set number[i] = choosing[i] = 0 and then we jump to label L1. Otherwise, we will continue by incrementing the maximum value and following the original Bakery algorithm.

We are more interested in proving that Bakery++ never allows an overflow and we will follow the style of structural proofs presented in [7]. We will also present two informal arguments about the properties that Bakery++ satisfies. We first need to define an overflow. Consider a computer with a set of identical registers and let M be the maximum value that may be stored in each register. We say that an overflow will occur if a value greater than M is to be stored in a register.

## 6.1  Proof of Avoiding Overflows

**Theorem**    Bakery++ never allows an overflow.

1. It suffices to assume
    1.1  M is a natural number representing the maximum value that a computer register may store.
    1.2  The processes follow the instructions of Bakery++ presented in Algorithm 2.
    1.3  An overflow occurs if a process tries to store a value $v$ in a register such that $v > M$.
    1.4  The registers of processes can hold a value $v$ such that $0 \leq v \leq M$.
    1.5  A process that crashes (halts) inside its critical section will go to its noncritical section and sets its shared variables *number* and *choosing* equal to 0.
    1.6  A process is allowed to fail at any instant and then to restart in its noncritical section.
    1.7  Any read operation that tries to read a memory location of a crashed process will return the value 0 eventually.

    and prove if there is a possibility of an overflow, then the corresponding register is set to 0.
    PROOF: The only instruction in Bakery that may lead to an overflow is handled in Bakery++.
2. The only possible instruction that may cause an overflow is "number[i] := number[i] + 1;".
    PROOF: By 1.3 and Algorithm 2, that is the only instruction that can store a value $v$ in a register such that $v \geq 1$.
3. In Bakery++, a value $v$ cannot be stored in a register such that $v > M$.
    PROOF: By 2 and Algorithm 2, each process will check the value of number[i] before incrementing it. If the value is too large, then the process will not increment it.
4. Q.E.D.
    PROOF: By 1.3 and 3.

## 6.2  Safety Argument

The mutual exclusion property, asserting that no two processes may enter their critical sections simultaneously, is a safety property. Bakery++ is almost identical to the Bakery algorithm and it maintains the mutual exclusion property. The basic changes that we made to Bakery are two simple conditional statements. We have not set values to variables in any way that differs from Bakery. We begin our argument by describing the first change that we made to Bakery. At the beginning of Bakery++, a process will check whether there is a possibility of an overflow in the near future. If there is a process that has stored the maximum possible value for a register, then there is a possibility of an overflow. We say this situation is *illegitimate*.

If a process detects that there is such a possibility by inspecting the values of all processes, then it waits until the situation becomes legitimate, meaning that the process waits until it reads a value less than the maximum possible value from all processes for their corresponding variable *number*. This check is performed at the beginning of each cycle of a process, therefore, all processes that have not crashed will perform the same check and *luckily,* all of them will make the same decision. We say luckily because it is possible for a read operation that occurs simultaneously with a write operation to return any arbitrary number!

We continue our argument by describing the second change that we have made to Bakery. What Bakery does is that it increments the value returned by the function *maximum* without any additional checks. We have added a conditional statement to check if the value to be incremented is too large. If the value is not too large, then we can do as Bakery does by incrementing the value and continuing the instructions. Otherwise, we set the values of variables *number* and *choosing* equal to 0 and then we jump to the beginning of the algorithm, which is label L1. There is another minor change that we have made to Bakery and that is defining a constant variable M that represents the maximum value that a register may contain.

We conclude our argument by claiming that we have not changed the execution flow of the Bakery algorithm in any specific way. If Bakery satisfies a property P, then Bakery++ satisfies it too. We have specified Bakery++ in the PlusCal algorithm language and we have model checked the specification using the TLC model checker for the properties of avoiding overflows and mutual exclusion that asserts no two processes ever enter their critical sections at the same time. In other words, if one wanted to write a proof that Bakery++ satisfies the property of mutual exclusion, then the same proofs that were written in [6] can be used again. Bakery++ can be thought of as a refinement of Bakery because every execution of Bakery++ is a valid execution of Bakery.

## 6.3 Liveness Argument

We study the possibility of processes being deadlocked while they are at label L1 in Algorithm 2. The correctness condition 2 given in Section 1.2 is a liveness property. It is worth mentioning that the Bakery algorithm does not guarantee liveness. If a process *j* keeps failing and restarting, then it is possible for process *i* to get stuck at label L2 [6]. Therefore, the same holds for Bakery++.

Our main concern in this section is whether processes are stuck at label L1. Let M be the maximum possible value that a register can store. The only possible situation in which this might happen is if a process reads M for the variable number of some process repeatedly. In theory, this is possible because we may have an extremely slow process against two processes that are quite fast. The slow process, namely k, may read the value M from one of two other processes, and then the two processes reset their registers to 0 but k is too slow to read the value 0 and the two fast processes keep competing and then they reach M again! Perhaps having such an incredibly slow process is equivalent to not having it in practice.

## 7 Performance and Practicality

Bakery++ does not introduce new variables and it manipulates the same data objects that Bakery does. Therefore, the spatial complexities of both algorithms are identical and it depends on the total number of processes. There are two arrays with size N and one variable *j* for each process, therefore, both Bakery and Bakery++ have *O(N)* spatial complexity.

The temporal complexity of Bakery++ depends on the number of executions of the *goto* statement exactly after label L1 in Algorithm 2. If there will not be an overflow, then Bakery++ has the same temporal complexity as Bakery because its control flow is almost identical to Bakery. If there are going to be many overflows and we have some extremely fast processes that can cause overflows while others are waiting for them, which is highly unlikely, then the temporal complexity of Bakery++ becomes more expensive compared to Bakery. This is the price of guaranteeing that no overflows ever occur.

The reason Bakery++ is useful in practice is that it is almost as simple as the original Bakery, without new variables and fancy operations or functions. Any existing implementation of Bakery can be extended to Bakery++ with the least effort due to its simplicity. We have leveraged two simple conditional statements to design Bakery++.

There are no practical limitations for implementing the Bakery++ algorithm. As a practical application of Bakery++, a multi-core modern laptop may implement it in order to guarantee that only a single thread in a group of threads can access a shared resource, such as a file.

## 8 Discussion

We put forward two questions about the Bakery++ algorithm that can serve as future study subjects.

### 8.1 Question One

What happens if there are more customers in the bakery than the maximum number that can be written on the tickets by the teller machine? This is quite similar to the question posed by Yoon et al. [12] and they pointed out that this is improbable in a real-world setting. The question is still of importance from a theoretical perspective. If there are more customers than the maximum value that may be written on a ticket, then is it possible that the Bakery algorithm will not be able to satisfy the correctness condition that asserts any process who wishes to enter its critical section should be allowed to do so eventually?

This problem arises from the possibility that a computer may have 8-bit registers (e.g., a tiny microprocessor) and yet we may be willing to implement the Bakery algorithm with 300 such processors. One possible solution may be compiler-based, i.e. the programmer may assume that only a single register is being used while in reality two or more registers are involved.

### 8.2 Question Two

The customers are served on a first-come-first-served basis in Bakery and Bakery++. Bakery had no trouble with arbitrary large ticket numbers because it assumed registers that could store infinitely large natural numbers. There is an upper bound in Bakery++ that limits the maximum possible value for a ticket. One may argue that we are storing the value of the *maximum* function in a variable and then checking the possibility of a future overflow, therefore, a process that has "*taken a turn*" to enter its critical section may not be able to use the given ticket because the value on the ticket is too large.

The truth of this statement depends on how we interpret the concept of "*taking a turn*" by a customer. Taking a turn may correspond to receiving a ticket (i.e. when the process sets the value of *number* equal to 1 + maximum (number[1], …, number[N])), or it may correspond to setting the value of *choosing* equal to 0. We argue that a process has not yet taken a turn in entering its critical section unless it increments the value that is returned by the *maximum* function. The question remains whether there is a firm explanation of which interpretation is true.

## 9 Conclusion and Future Work

Whenever there is a need to share a resource, mutual exclusion algorithms come to help by providing users with a coordination scheme. We revisited the problem of mutual exclusion and the Bakery algorithm, the first true mutual exclusion algorithm, to solve the problem of integer overflow in the original Bakery.

We presented a modified version of the Bakery algorithm, named Bakery++, and we proved that Bakery++ never allows an overflow. Bakery++ solves the problem of register overflow without introducing new variables or redefining the operations or functions in the Bakery algorithm. Bakery++ is almost as simple as Bakery and this allows straightforward implementations of Bakery++ in real systems. Bakery++ facilitates the management of access to resources in a parallel system without the possibility of overflows occurring, and it does so in a straightforward manner with the least modifications to the original Bakery algorithm.

Previous approaches to solving the problem of register overflow concentrated on making use of new variables and redefining the operations or functions used in the original Bakery, and they were somehow complicated solutions. Bakery++ is quite simple and it differs from Bakery in just a few instructions. We presented arguments about safety and liveness of Bakery++, bearing in mind that Bakery++ is slightly different from Bakery. We showed that Bakery++ has the same temporal and spatial complexities as the original Bakery algorithm and we highlighted the practicality of Bakery++ due to the elimination of overflow possibilities. We specified Bakery++ in PlusCal language and we performed model checking on the specification to assert that Bakery++ satisfies mutual exclusion and never allows overflows.

The new algorithm, namely Bakery++, is formally specified in the PlusCal language and verified correct using the TLC model checker. The problem of register overflow is important because if an overflow occurs, then the correctness of the system will be lost. Bakery++ solves the register overflow problem differently from all existing solutions. Bakery++ does not use additional memory units and it does not modify the definition of functions or arithmetic operations used in the original Bakery algorithm. Also in Bakery++, the processes are not allowed to write into other processes' memory locations. Therefore, Bakery++ is original and optimal.

We expressed two open questions that are of interest in the context of Bakery++ and they may act as basic problems for future studies. The first question considers the scenario in which there are too many customers in the bakery who wish to be served and we need to know if Bakery++ can satisfy the Correctness Condition 2 given in Section 1.2. The second question is subjective and it is about giving a precise definition of the exact moment when a process is considered to have taken its ticket from the teller machine. The receipt of a ticket means that the customer should be served before all those who will request service later.

## REFERENCES

[1] Alex A. Aravind. 2010. Highly-Fair Bakery Algorithm Using Symmetric Tokens. *Inf. Process. Lett.* 110, 23 (2010), 1055–1060. DOI:https://doi.org/10.1016/j.ipl.2010.09.004
[2] Alex A. Aravind and Wim H. Hesselink. 2011. Nonatomic Dual Bakery Algorithm with Bounded Tokens. *Acta Inform.* 48, 2 (2011), 67–96. DOI:https://doi.org/10.1007/s00236-011-0132-0
[3] Edsger Wybe Dijkstra. 1965. Solution of a Problem in Concurrent Programming Control. *Commun. ACM* 8, 9 (1965), 569. DOI:https://doi.org/10.1145/365559.365617
[4] Prasad Jayanti, King Tan, Gregory Friedland, and Amir Katz. 2001. Bounding Lamport's Bakery Algorithm. In *Proceedings of the International Conference on Current Trends in Theory and Practice of Computer Science*, Springer, Piešt'any, Slovak Republic, 261–270. DOI:https://doi.org/10.1007/3-540-45627-9_23
[5] Donald E. Knuth. 1966. Additional Comments on a Problem in Concurrent Programming Control. *Commun. ACM* 9, 5 (1966), 321–322. DOI:https://doi.org/10.1145/355592.365595
[6] Leslie Lamport. 1974. A New Solution of Dijkstra's Concurrent Programming Problem. *Commun. ACM* 17, 8 (1974), 453–455. DOI:https://doi.org/10.1145/361082.361093
[7] Leslie Lamport. 2012. How to Write a 21st Century Proof. *Journal of Fixed Point Theory and Applications* 11, 1 (2012), 43–63. DOI:https://doi.org/10.1007/s11784-012-0071-6
[8] Leslie Lamport. The Writings of Leslie Lamport. Retrieved March 6, 2020 from http://lamport.azurewebsites.net/pubs/pubs.html
[9] F. W. van der Sommen, W. H. J. Feijen, and A. J. M. van Gasteren. 1997. Peterson's Mutual Exclusion Algorithm Revisited. *Science of Computer Programming* 29, 3 (1997), 327–334. DOI:https://doi.org/10.1016/S0167-6423(97)00003-8
[10] Boleslaw K. Szymanski. 1990. Mutual Exclusion Revisited. In *Proceedings of the Fifth Jerusalem Conference on Information Technology*, IEEE, Jerusalem, Israel, 110–117. DOI:https://doi.org/10.1109/JCIT.1990.128275
[11] Gadi Taubenfeld. 2004. The Black-White Bakery Algorithm and Related Bounded-Space, Adaptive, Local-Spinning and FIFO Algorithms. In *Proceedings of the International Symposium on Distributed Computing* (DISC '04), Springer, Amsterdam, The Netherlands, 56–70. DOI:https://doi.org/10.1007/978-3-540-30186-8_5
[12] Dong Young Yoon, Mosharaf Chowdhury, and Barzan Mozafari. 2018. Distributed Lock Management with RDMA: Decentralization without Starvation. In *Proceedings of the International Conference on Management of Data* (SIGMOD '18), ACM, Houston, TX, USA, 1571–1586. DOI:https://doi.org/10.1145/3183713.3196890